\documentclass[9pt,a4paper]{scrartcl}
\usepackage{helvet}
\usepackage[a4paper, lmargin=1.7cm, rmargin=1.7cm, tmargin=3cm, bmargin=3.2cm]{geometry}

 \usepackage[utf8x]{inputenc}
\usepackage[T1]{fontenc}

\usepackage{amsmath}
\usepackage{bm}
\usepackage{upgreek}

\usepackage{hyperref}

\usepackage{pdfpages}
\usepackage{graphicx}

\begin{document}



# Experiment-friendly kinetic analysis of single molecule data in and out of equilibrium

Sonja Schmid[a], Markus Götz[a], Thorsten Hugel[a,1]

[a]Institute of Physical Chemistry, University of Freiburg, Albertstrasse 23a, 79104 Freiburg, Germany; [1]thorsten.hugel@pc.uni-freiburg.de

We present a simple and robust technique to extract kinetic rate models and thermodynamic quantities from single molecule time traces. SMACKS (Single Molecule Analysis of Complex Kinetic Sequences) is a maximum likelihood approach that works equally well for long trajectories as for a set of short ones. It resolves all statistically relevant rates and also their uncertainties. This is achieved by optimizing one global kinetic model based on the complete dataset, while allowing for experimental variations between individual trajectories. In particular, neither *a priori* models nor equilibrium have to be assumed. The power of SMACKS is demonstrated on the kinetics of the multi-domain protein Hsp90 measured by smFRET (single molecule Förster resonance energy transfer). Experiments in and out of equilibrium are analyzed and compared to simulations, shedding new light on the role of Hsp90's ATPase function. SMACKS pushes the boundaries of single molecule kinetics far beyond current methods.

molecular machines | conformational kinetics | energy conversion | single molecule | FRET

The ability to reveal conformational state sequences at steady state is a unique feature of single molecule time traces. Conformational kinetics is detectable in or out of equilibrium, which enables direct calculation of thermodynamic quantities. Single molecule Förster resonance energy transfer (smFRET) is one of the most common methods to do so. According to the current standard analysis of kinetic smFRET trajectories, state sequences are deduced using hidden Markov models (HMM) (1–3) and rates are then obtained from single-exponential fits to the respective dwell time histogram of every observed state.

This standard approach is feasible under the following two conditions: First, every state has a characteristic FRET efficiency. Second, all transition rates are similar. In this case, there is a sampling rate at which every state is reached many times before irreversible photo-bleaching. Both requirements are broken by regular proteins, which commonly exhibit rates on diverse time-scales and conformations that are experimentally indistinguishable, but differ kinetically (kinetic heterogeneity) (4–6). As a consequence, multi-exponential dwell time distributions are obtained. The interpretation of such distributions may lead to erroneous conclusions (see below).

With our new experiment-friendly approach, we overcome these problems by training one global HMM based on a set of experimental time traces. The procedure copes with experimental shortcomings and kinetic heterogeneity. Further, it provides several means of model evaluation including error quantification. Finally, we demonstrate how to deduce kinetics and thermodynamics of the heat-shock protein Hsp90.

## Results

**Rate extraction from an ideal model system.** Holliday-junctions (7) have become a widely used model system for conformational dynamics studied by smFRET. These DNA four-way junctions alternate constantly between two equilibrium conformations (8). Such dynamics were recorded by a custom-built objective-type total internal reflection fluorescence (TIRF) microscope (Fig. 1*A*) with alternating laser excitation (ALEX) (9). An example trace is shown in Fig. 1*B*. As expected for a two-state system, the FRET histogram shows two peaks (Fig. 1*C*) and the dwell time histograms are well fit by single-exponential functions (Fig. 1*D*). In this case, all standard methods work well and the extracted rates will be correct.

**Rate extraction from typical protein systems.** In contrast, the situation is more complicated for proteins, which usually adopt significantly more than two states (10). As an example, we show equivalent single protein time traces revealing conformational changes of the heat-shock protein Hsp90 (11) (Fig. 1*E*). This homo-dimeric protein fluctuates between N-terminally open and closed conformations (12) resulting in two peaks in the FRET histogram (Fig. 1*F*). The fluctuations occur on a broad range of time-scales resulting in very long and short dwells, and generally fewer transitions per trace (here 3 on average). Despite the two apparent FRET populations, both dwell time distributions are multi-exponential (Fig. 1*G*). Yet, no systematical change in FRET efficiency from fast to slow dwells is observed (Fig. S1). Such behavior (hereafter referred to as *degenerate FRET efficiencies*) is indicative of truly hidden states that cannot be separated by FRET efficiency, but differ kinetically.

The kinetic analysis is complicated by the limited detection bandwidth of smTIRF experiments. It is restricted by the exposure time, on the one side, and the mean observation time - limited by photo-bleaching - on the other side. While enzymatic anti-bleaching agents largely increase the observation time of DNA-based samples, those are much less effectual against bleaching of all-protein systems. Furthermore, their use with protein systems is problematic, as they might interact with the protein under study. Accordingly, the detection bandwidth spans less than a factor of 200 at a reasonable signal to noise ratio – independent of the sampling rate applied.

In this situation, the classical dwell time analysis ignores large parts of the data, because only dwells with clearly defined start and end points are considered. As a consequence, predominantly long dwell times are missed, resulting in transition rates that are systematically overestimated. Already in a two state system, deviations of more than a factor of two occur (Fig. 2*A*). Importantly, even so-called static traces (without any transition) contain kinetic information. They occur in the experiment as a result of the finite observation time, especially if both fast and slow processes occur. In other words: the presence of at least two transitions per trace is an inappropriate and misleading criterion for trace selection. Yet it is the intrinsic requirement of dwell time analysis. Moreover, the connectivity of states is completely ignored by dwell time analysis. Please note that the limitations of dwell time analysis have been recognized in the patch clamp field more than 20 years ago (13). Nevertheless, it is still the standard analysis in the smFRET field today.

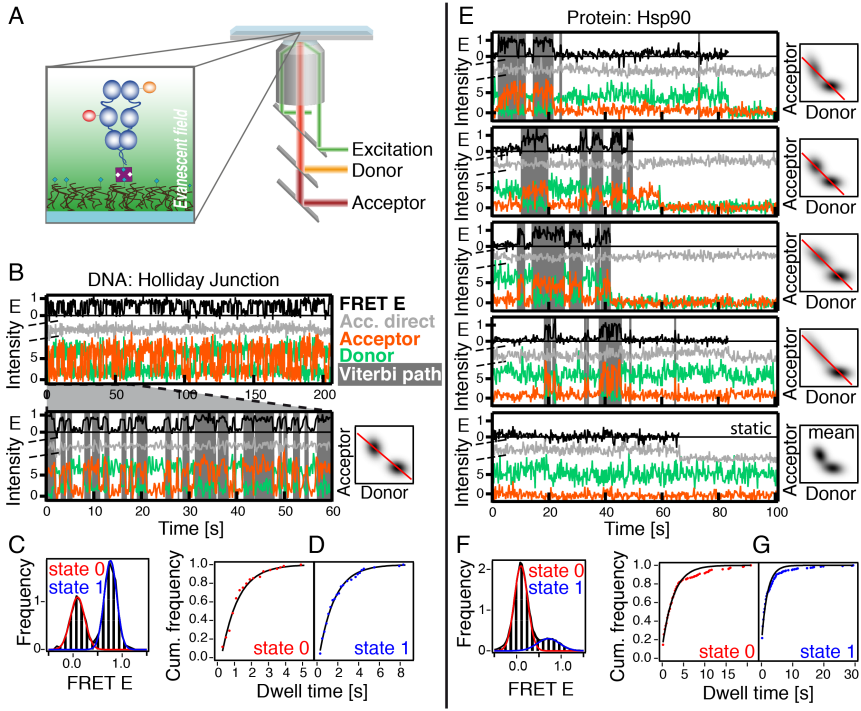

**Fig. 1.** Conformational dynamics of a DNA (left) and a protein system (right) measured by smFRET. (a) Schematic TIRF setup with ALEX and close-up of a single Hsp90 molecule, labeled with a FRET pair (donor dye orange, acceptor red) and immobilized within the evanescent field (see Methods). (b, e) FRET efficiency (E) trajectories (black) obtained from single molecule fluorescence time traces (donor green, acceptor orange, in arbitrary units). Fluorescence of the acceptor dye after direct laser excitation (gray) excludes photo-blinking artifacts. The HMM derived state sequence (Viterbi path) is displayed as overlays (low FRET white, high FRET gray). A zoom is included in (b). Emission PDFs in dimensions of donor and acceptor fluorescence are shown next to the corresponding time traces. The allowed "FRET line" (red) is determined for every molecule individually (cf. main text). (c, f) FRET histograms with Gaussian fits as indicated. (d, g) Cumulative dwell time histograms with single-exponential fits (black, frequency weighted). Static traces (e, bottom) are described using the mean emission PDFs of the complete dataset (cf. main text). Although the FRET histogram (f) shows two populations, the dwell time distributions (g) clearly deviate from a single-exponential fit (total of 163 molecules).

**A better solution for typical protein systems.** In view of the experimental reality, we developed a new Single-Molecule Analysis for Complex Kinetic Sequences – short: SMACKS. It combines all experimentally available information in one HMM, which allows us to investigate important thermodynamic concepts that go significantly beyond dwell time analysis.

Such an HMM consists of invisible or "hidden" kinetic states that generate certain detectable signals (e.g. high FRET, low FRET) with a given probability. The sequence of states is assumed to be memory-less, i.e. the probability of a certain transition depends only on the current state. Any time-homogeneous Markovian analysis requires stationarity – but *not* thermodynamic equilibrium. An HMM is parameterized by one start probability $\pi_i$ per state $i$, transition probabilities $a_{ij}$ between all hidden states assembled in the transition matrix $A$, and a set $B$ of so-called emission probabilities $b_i$ that link the hidden states to the observables (14, 15).

By exploiting the original *two* observables - donor and acceptor fluorescence - instead of the FRET efficiency (only *one* observable), the robustness with respect to uncorrelated noise is significantly increased. These fluorescence signals are appropriately described by 2D Gaussian probability density functions (PDFs), $b_i(\vec{\mu}, V)$, parameterized by the means $\mu$ and the co-variance matrix $V$ - all in dimensions of donor and acceptor fluorescence. Representative emission PDFs are graphed at the right hand side of Fig. 1B,E.

The mathematically available parameter space for emission probabilities is further restricted by physical knowledge about FRET. Namely, the mean total fluorescence intensity is required to remain constant within one trace (Eq. 1 and *SI Methods*), whereas experimental variations between individual molecules are tolerated.

$$\langle I_{tot} \rangle = \mu_D + \mu_A = \text{const.} \quad [1]$$

Here the donor and acceptor intensities (with means $\mu_D$ and $\mu_A$) were corrected for background, experimental cross-talk and the gamma factor beforehand. The resulting allowed "FRET-line" is displayed in the emission graphs (Fig. 1B,E right).

In the classical HMM implementation (14), the model $\lambda(\pi, A, B)$ is iteratively rated by the forward-backward algorithm and optimized by the Baum-Welch algorithm until convergence to maximum likelihood. The Viterbi algorithm is used to compute the most probable state sequence for every trace given the previously trained model. In contrast to earlier published ensemble approaches (16–19), SMACKS works without additional (hyper-) parameters or prior discretization.

The full procedure was tested on various synthetic datasets generated by known input models, in or out of equilibrium, with or without degenerate FRET efficiencies. Synthetic data contained noise, photo-bleaching, randomly offset individual traces and a realistic dataset size (see *SI Methods* and example data in Fig. S2).

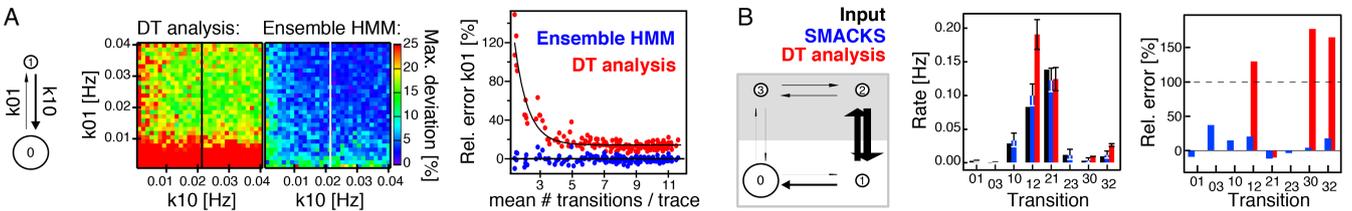

**Fig. 2.** Accuracy of rates from dwell time analysis compared to ensemble HMM. (a) Discrete state sequences were simulated for 2-state models (left) with different rates, k01, k10 (200 state sequences with 5Hz sampling rate and 0.03Hz bleach rate). The deviation of the determined k01 from the input k01, as obtained by dwell time (DT) analysis and ensemble HMM, is shown as a function of both input rates (maximum relative deviation out of 5 simulations per data point). Relative errors of the rates along the indicated lines are shown as a function of the mean number of transitions per trace (right). Black lines serve as a guide to the eye. Clear systematic deviations occur already for the simplest model. (b) More complex models are more realistic for protein systems. The depicted 4-state model (left) was used to generate synthetic data with equal FRET efficiencies for states 0/1 and 2/3, respectively. See *SI Methods* for details and Fig. S2 for example data. The size of the circles in the state models (a and b, left) is proportional to the state population. The arrow widths are proportional to the transition rates. Right: comparison of the transition rates determined by DT analysis and SMACKS. DT analysis only provides half the transition rates, which are in addition far off the input values.

2

SMACKS resolved accurate transition rates despite degenerate FRET efficiencies, where neither dwell time derived rates nor error estimates were meaningful (Fig. 2B).

**Demonstration of SMACKS using experimental data.** We start with a set of smFRET time traces obtained with alternating laser excitation (ALEX) that were selected and corrected as previously described(20) (*SI Methods*). Namely, the donor and acceptor intensities ($I_D$, $I_A$) satisfy Eq. **2**. However, previous smoothing is not required.

$$\text{FRET E} = \frac{I_A}{I_D + I_A} \quad [2]$$

An *apparent* state model can be deduced from visual inspection of the FRET time traces and the FRET histogram. This model (2 states for Hsp90) is used in a first trace-by-trace HMM optimization to train individual emission PDFs on each molecule separately. The trained parameters are examined visually by comparing the resulting Viterbi path to the input data. Notably, by searching for flat plateaus, HMM echoes a characteristic requirement for single-molecule fluorescence data.

For static traces (here 34% of all traces), a model with more than one state will not converge sensibly. Therefore, static traces are included using the mean emission PDFs of the remaining dataset (see fifth molecule in Fig. 1E).

As a next step (Fig. 3A), an *ensemble* HMM run is performed to optimize the start and transition probabilities based on the *entire* dataset, while holding the predetermined, individual emission PDFs fixed. While different strategies have been tested, this solution worked equally well for experimental *and* simulated data. The kinetic heterogeneity found in Hsp90 is investigated by comparing different state models including duplicates and triplicates of the *apparent* states (Fig. 3B).

Similar to others (1, 3, 21), we then use the Bayesian information criterion (BIC) (22) for model selection (see *SI Methods*). We find that Hsp90's conformational dynamics are best described by a 4-state model with 2 high FRET (closed) and 2 low FRET (open) states. This is consistent with the bi-exponential dwell time distributions shown in Fig. 1G.

Once the optimal number of states is deduced, the model is further refined by inspecting the Viterbi paths. The transition map (Fig. 3C left) shows the quality of both, the original input data and the state allocation based on the obtained model. It reveals the clustering of the transitions in FRET space. Importantly, the transition map itself cannot report on the number of states in the model, because it is the consequence of a predetermined model. The occurrence of all transitions is shown in a 2D histogram (Fig. 3C middle). For a system functioning at thermodynamic equilibrium, detailed balance requires that the transition histogram is symmetric about the main diagonal. Out of 12 possible transitions in a fully connected 4-state model, only 8 cyclic transitions are populated for Hsp90 with ATP. Despite the reduced number of free parameters, a cyclic 4-state model fits the data with equal likelihood. This is in line with the maximal number of theoretically identifiable transitions (8 in the case of 2 open (o) and 2 closed (c) states (23)). While being difficult to interpret in the context of Hsp90, a cyclic -o-c-o-c- model would theoretically fit the data equally well. Further information on the interpretation of degenerate state models is given in (23–25).

**Model evaluation.** In most previous kinetic studies on smFRET, the only reported error estimates were the uncertainties of fit coefficients from fitting dwell time distributions, disregarding systematic overestimation and variations throughout the dataset (Fig. 2). In contrast, we propose three tests to assess the reliability of the results from the above procedure.

First, the most illustrative test for the consistency of the trained model with the original data is "re-simulation" using the obtained transition matrix, the experimental bleach rate and degenerate states (here 2o, 2c). Fig. 3D (left) shows very good agreement between the re-simulated and the experimental dwell time distribution. FRET histograms can be re-simulated, too.

Second, the convergence of the HMM to the global maximum is tested by using multiple random start parameters (26). In all attempts, the parameters converged to the same maximum likelihood

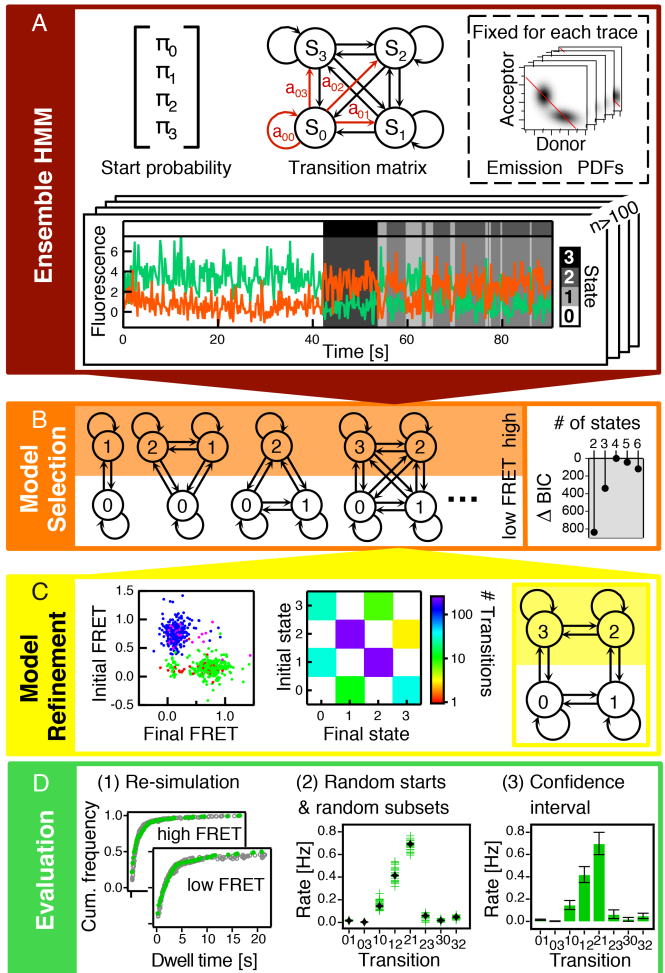

**Fig. 3.** SMACKS workflow. (a) The model parameters (start probability $\pi$, transition matrix $A$, emission PDFs $B$) and a set of donor (green) and acceptor (orange) fluorescence time traces constitute the basis of ensemble HMM (overlays represent the Viterbi path). To allow for inter-molecule variations of the fluorescence signal, the individual emission PDFs are predetermined in a trace-by-trace HMM run and held fixed during optimization of the kinetic ensemble parameters. (b) Degenerate FRET states are included as multiples of experimentally discernible states (here low FRET and high FRET). The Bayesian information criterion (BIC) identifies the optimal model (here 4 states). (c) The transition map (left) relates the mean FRET values before and after each dwell found by the Viterbi algorithm (initial state 0, 1, 2, 3 in red, green, blue, pink, respectively). The *most frequent* transitions occur between the two *least populated* (short-lived) states. The transition histogram (middle) reveals the frequency of each transition in the dataset. Excluding transitions that do not occur leads to a cyclic model (right). (d) The obtained model is critically evaluated in three ways: First, dwell time distributions are reproduced by re-simulating the model (left, experimental data: green, simulated: gray). Second, random start parameters uncover potential local likelihood maxima, and random subsets reveal dataset heterogeneity (middle, subsets: green, complete set: black). Third, confidence intervals measure the precision of the obtained rates considering the finite dataset, experimental noise and dataset heterogeneity (right). A flow chart of SMACKS can be found in Fig. S3.

3

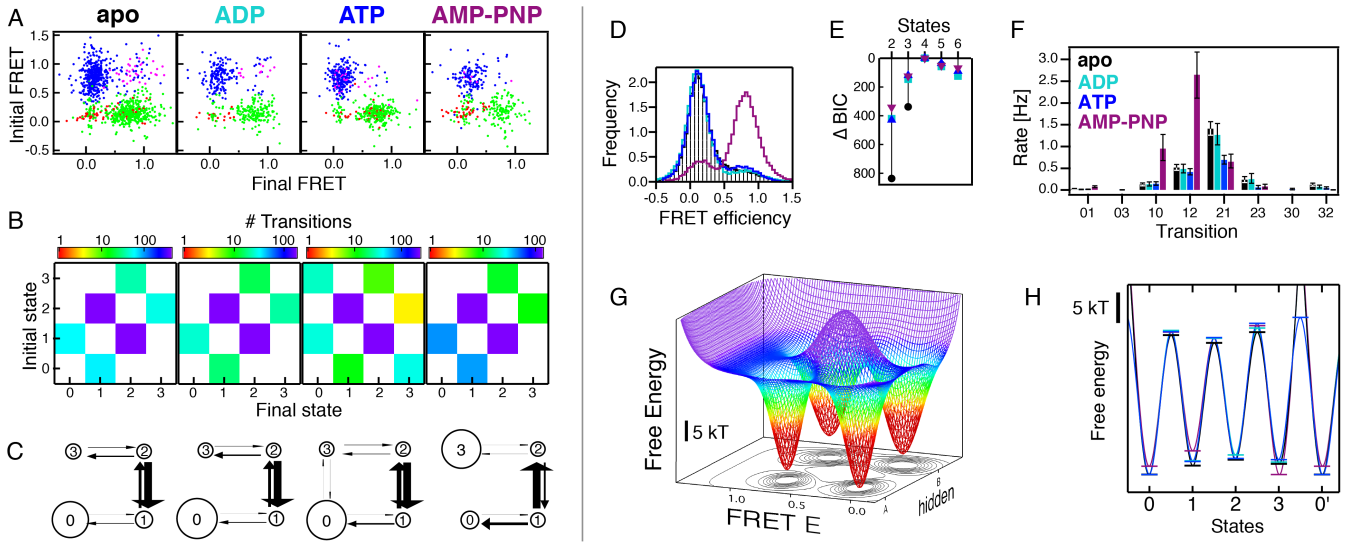

**Fig. 4.** Kinetic and thermodynamic results under varied nucleotide conditions. (a) Transition maps locate transitions in FRET space (initial state 0, 1, 2, 3, in red, green, blue, pink). FRET E scatters less in the presence of nucleotides. (b) Occurrence of transitions. (c) Derived kinetic state models: states 0/1 (2/3) show low (high) FRET efficiencies, respectively. Arrow widths represent the size of the transition rates. Circle sizes represent relative state populations. A 4-link model is favored with ATP, while 3 links are sufficient for all other datasets. (Rates to remain in a given state are not depicted.) (d) The FRET histogram shows only small differences for apo (black and shaded black), ADP or ATP data, whereas with AMP-PNP (purple) a shift to the closed conformation is observed. (e) A 4 state model represents all four datasets best according to ΔBIC values, color code as in (f). (f) Deduced rates and confidence intervals. (g) Qualitative cartoon of the 3D energy surface of Hsp90 in the presence of ATP. SMACKS reveals "hidden" states that are kinetically different while sharing the same FRET efficiency. A large energy barrier hinders transitions through the midpoint. (h) Quantitative 2D projection of the energy landscape shows the differences between the nucleotides, color code as in (f). Energy levels were calculated from transition rates, whereas well widths are arbitrary. A typical attempt frequency for proteins, $10^8$Hz, was assumed.

estimators (MLE). Additionally, random subsets of the data (here 66% of the traces) reveal the heterogeneity of the dataset (Fig. 3*D* middle). Here all random subsets (n > 50) converged to the same model with normally distributed parameters (RMSD ≈ 30% of the MLE).

Third, the confidence interval of every trained parameter is calculated using likelihood ratio tests (Fig. 3*D* right) (3, 27). It reports on the dataset heterogeneity and the precision of the HMM (see *SI Methods*).

In summary, the accuracy and precision of the kinetic state model deduced by our semi-ensemble HMM approach was demonstrated by threefold evaluation. A reliable state model is necessary to take the next step and resolve kinetic and thermodynamic information from proteins in or out of equilibrium.

**The kinetic model of Hsp90.** Hsp90 is a slow ATPase (28) and changes between open and closed conformations at room temperature. In Fig. 4, kinetic results for Hsp90's conformational dynamics are compared under different nucleotide conditions (2mM ADP, ATP, AMP-PNP) or without nucleotides (apo). The quality of the input data and the resulting state allocation is visible on the transition map in FRET space (Fig 4*A*). It is evident that Hsp90's conformational changes are less defined in the absence of nucleotides. The same FRET efficiencies are detected in all experiments: $E_{low}$=0.1, $E_{high}$=0.8 (Fig. 4*D*). Consistently, the transitions cluster around these efficiencies. The optimal state model contains four states under all conditions (Fig. 4*E*): states 0/1 are long-/short-lived low FRET states, states 2/3 are short-/long-lived high FRET states, respectively (Fig. 4*C*). While four links are required to describe Hsp90's conformational dynamics in the presence of ATP (Fig. 4*B,C*), under apo, ADP and AMP-PNP conditions models with three links are sufficient (detailed in *SI Note 1*) (23). The rates under apo and ADP conditions are similar to those in the presence of ATP (Fig. 4*F*). Only with the non-hydrolysable ATP-analogue, AMP-PNP, the rates between both short-lived states are inverted.

This is in agreement with the pronounced shift towards the closed conformation observed in the FRET histogram, in the presence of AMP-PNP (Fig. 4*D*).

**Exploring energy coupling.** Protein machines, such as Hsp90, use external energy (e.g. from ATP hydrolysis) and therefore operate out of equilibrium. A central question is where (in the conformational cycle) energy consumption couples into protein function. Based on SMACKS, we can address this question quantitatively. It boils down to determining the free energy difference over closed cycles (29, 30) (in units of thermal energy, kT):

$$\Delta G_{cyc} = - \sum_{\substack{\forall i \neq j \\ \text{(cycl.)}}} \ln\left(\frac{a_{ij}}{a_{ji}}\right) \quad [3]$$

As expected and required, we find that in the absence of an external energy source, Hsp90's conformational dynamics are at equilibrium. At first sight unexpected, we find for Hsp90 in the presence of ATP $\Delta G_{cyc} = (0.9 \pm 0.9)$ kT. This indicates that the energy of ATP hydrolysis is not coupled to the observed conformational changes, which is consistent with earlier results (12). A schematic energy landscape is shown in Fig. 4*G*. The 3D illustration highlights SMACKS ability to split two observable FRET states into four states based on their distinct kinetic behavior. Quantitative energies are shown in Fig. 4*H*.

**Experimental limits for resolving energy coupling.** Clearly, the accuracy of the resolved $\Delta G_{cyc}$ depends on the size of the dataset. Especially for systems away from equilibrium, very slow ("reverse") rates can occur. Due to the finite dataset, only few respective transitions are observed, resulting in large relative errors for these small rates. In this case, an alternative formulation of $\Delta G_{cyc}$ using the number of transitions $N^{\text{trans.}}$ found by the Viterbi algorithm is more robust:

$$\Delta G_{cyc} \approx - \sum_{\substack{\forall\, i \neq j \\ (\text{cycl.})}} \ln\left(\frac{N_{ij}^{\text{trans.}}}{N_{ji}^{\text{trans.}}}\right) \quad [4]$$

Eq. **4** represents a lower bound for the free energy difference, given the finite dataset (zero transitions are set to one to avoid poles). If all rates are well resolved, Eqs. **3** and **4** yield the same result.

In the following, two limit cases for the coupling of conformational changes to ATP hydrolysis are considered systematically ($\Delta G_{cyc}$ = 30 kT for ATP to ADP hydrolysis assuming 1% ADP, 3mM $Mg^{2+}$, 250mM KCl and 100% efficiency) (31). In the first case (Fig. 5*A*), the full 30 kT are introduced within one step. Whereas in the second case (Fig. 5*B*), the energy is successively released over four steps, comparable to contributions by ATP binding, hydrolysis and ADP or $P_i$ release, proposed e.g. for the human mitochondrial $F_1$-ATPase (32). Although realistic mechanisms will be a mixture of the two, these ideal cases allow for a systematic calculation of the maximally observable free energy $\Delta G_{obs}$ as a function of the dominating forward rate (Fig. 5*A,B* bottom). Even in the absence of noise and degenerate states, the observed free energy difference is limited by the finite dataset size. The same is true for the more realistic model shown in Fig. 5*C*: Eq. **4** applied to discrete state sequences yields 20.5 kT of the original 30 kT. This is because very unlikely transitions do not occur throughout the dataset (Fig. 5*C* bottom). After including all the experimental shortcomings and degenerate FRET efficiencies, SMACKS recovered $\Delta G_{cyc}$ = (12 ± 2) kT. This is 58% of the free energy, which was actually present in the synthetic data.

In view of these results, we stimulated Hsp90's hydrolysis rate more than tenfold by its co-chaperone Aha1 (33). If we had missed out on the directionality due to the slow ATPase rate, this should ultimately allow us to resolve putative energy coupling. Fig. 5*D* shows that even highly stimulated hydrolysis does not induce conformational directionality in Hsp90: $\Delta G_{cyc}$ = (−0.4 ± 1.2) kT in the presence of 3.5 μM Aha1. Our results strengthen the notion that Hsp90's large conformational changes are mainly independent of ATP hydrolysis.

**Discussion**

SMACKS is a novel HMM approach, which resolves all relevant rates that characterize the observed conformational dynamics, from a set of (short) smFRET time traces. The underlying states are identified by their FRET efficiency or kinetic behavior or both. SMACKS is a tailor-made solution for the wide family of protein machines that are clearly more challenging than DNA prime examples. It represents a significant advance that enables direct quantification of the energy coupled to conformational changes. This progress is achieved by the following six key features:

(i) SMACKS exploits the original fluorescence signal of the FRET donor and acceptor as 2D input. The FRET-specific anti-correlation provides significantly increased robustness with respect to uncorrelated noise. This unique information is lost in 1D FRET trajectories.

(ii) SMACKS tolerates experimental intensity variations between individual molecules, while at the same time, the transition rates are extracted from the entire dataset.

(iii) SMACKS minimizes the bias of photo-bleaching, because it determines transition rates based on their occurrence in the dataset. Thus, the range of detectable timescales can be expanded by increasing the dataset.

(iv) SMACKS performs the entire analysis on the experimental (i.e. noisy) fluorescence data. In fact, the knowledge about a given data point's reliability is used to weight its contribution accordingly. Therefore, SMACKS is robust enough to handle realistic noise levels in protein systems.

(v) SMACKS identifies hidden states that share indistinguishable FRET efficiencies, but differ kinetically.

(vi) SMACKS quantifies the precision of extracted rates. The precision is limited by the dataset size and signal quality, but it is not compromised by systematic overestimation, which contrasts with earlier studies.

The ATP-dependent molecular machine Hsp90 served here as an illustrative test case. SMACKS shed new light on the enigmatic and controversially discussed ATPase function (34). Clearly, the N-terminal conformational dynamics are not coupled to ATP hydrolysis, even in the presence of the co-chaperone Aha1. Further reaction coordinates will be explored by SMACKS to elucidate driven conformational changes and finally uncover the role of Hsp90's slow ATPase function.

In summary, our results demonstrate how SMACKS provides new power and confidence for the kinetic analysis of single molecule time traces in general. In particular for smFRET studies on sophisticated protein systems, SMACKS is unparalleled. We anticipate that SMACKS will reveal drive mechanisms in a large number of protein machines.

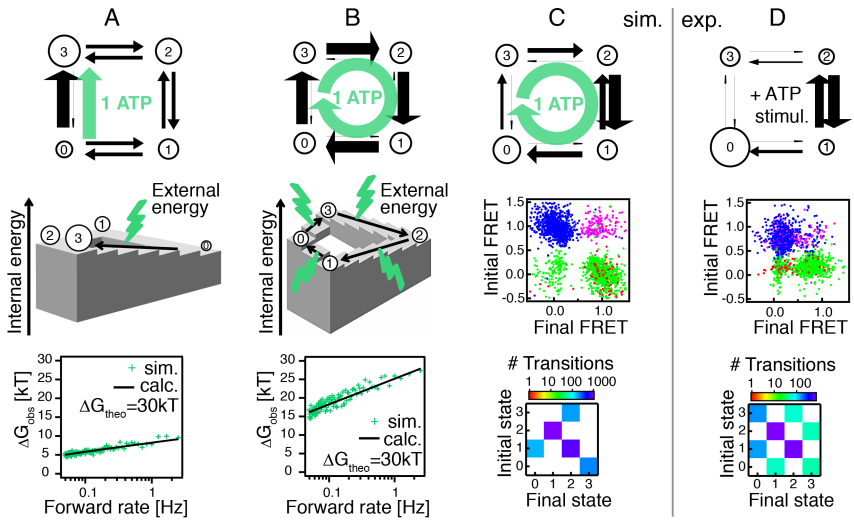

**Fig. 5.** Quantifying energy coupling. (a, b) Two limit cases of systems driven by the hydrolysis of 1ATP: in (a) the external energy is absorbed between states 0 and 3. All remaining rates are set to 0.05Hz. In (b) the external energy is introduced sequentially over 4 identical steps. Respective state models (top), energy scheme (center), and theoretical detection limit for free energies as a function of the forward rate (bottom). Simulated values (green) result from Eq. **4** applied to 200 discrete state sequences with 5Hz sampling rate and 0.03Hz bleach rate. They scatter about the expectation value of $\Delta G_{obs}$ (black line) calculated as explained in the *SI Note 2*. (c, d) State model (top), transition map (center) and transition histogram (bottom) for synthetic data (c) simulating the flow introduced by coupling to the hydrolysis of 1ATP = 30kT, or for experimental data (d) of Hsp90+ATP stimulated by the co-chaperone Aha1.

## Methods

Hsp90 or Holliday junctions specifically biotinylated and labeled with fluorescent dyes (Atto550/Atto647N maleimide) were immobilized on a passivated and Neutravidin coated fused silica coverslip that shows no auto-fluorescence upon ALEX (532nm or 635nm) in TIRF geometry using an EMCCD for detection. Measurements were performed at 5Hz at 21°C. More detailed descriptions are given in *SI Methods*. Monte Carlo simulations and HMM calculations were run in Igor Pro (Wavemetrics) on an ordinary desktop PC. Synthetic data contained Gaussian noise ($\sigma=0.3*signal$), random offsets ($\pm 0.2*signal$), degenerate FRET, efficiencies (two low / two high), a sampling rate of 5Hz and a bleach rate of 0.03Hz (see Fig. S2). All formulae utilized in semi-ensemble HMM (Forward-Backward, Baum-Welch and Viterbi algorithms) with continuous observables in 2D are included in *SI Methods*. The complete source code together with example data will be available shortly after publication at: http://www.singlemolecule.uni-freiburg.de/SMACKS .

1. McKinney SA, Joo C & Ha T (2006) Analysis of Single-Molecule FRET Trajectories Using Hidden Markov Modeling. *Biophys J* 91(5):1941-1951.
2. Bronson JE, Fei J, Hofman JM, Gonzalez RL Jr & Wiggins CH (2009) Learning Rates and States from Biophysical Time Series: A Bayesian Approach to Model Selection and Single-Molecule FRET Data. *Biophys J* 97(12):3196-3205.
3. Greenfeld M, Pavlichin DS, Mabuchi H & Herschlag D (2012) Single Molecule Analysis Research Tool (SMART): An Integrated Approach for Analyzing Single Molecule Data. *PLoS ONE* 7(2):e30024.
4. Frauenfelder H, Sligar S & Wolynes P (1991) The energy landscapes and motions of proteins. *Science* 254(5038):1598-1603.
5. Sunney Xie X (2002) Single-molecule approach to dispersed kinetics and dynamic disorder: Probing conformational fluctuation and enzymatic dynamics. *J Chem Phys* 117(24):11024-11032.
6. Henzler-Wildman K & Kern D (2007) Dynamic personalities of proteins. *Nature* 450(7172):964-972.
7. Holliday R (1964) A mechanism for gene conversion in fungi. *Genetics Research* 5:282-304.
8. McKinney SA, Declais A, Lilley DMJ & Ha T (2003) Structural dynamics of individual Holliday junctions. *Nat Struct Mol Biol* 10(2):93-97.
9. Kapanidis AN et al. (2005) Alternating-Laser Excitation of Single Molecules. *Acc Chem Res* 38(7):523-533.
10. Ansari A et al. (1985) Protein states and proteinquakes. *Proc Natl Acad Sci USA* 82(15):5000-5004.
11. Ali M et al. (2006) Crystal structure of an Hsp90-nucleotide-p23/Sba1 closed chaperone complex. *Nature* 440(7087):1013-1017.
12. Mickler M, Hessling M, Ratzke C, Buchner J & Hugel T (2009) The large conformational changes of Hsp90 are only weakly coupled to ATP hydrolysis. *Nat Struct Mol Biol* 16(3):281-286.
13. Horn R & Lange K (1983) Estimating kinetic constants from single channel data. *Biophys J* 43(2):207 - 223.
14. Rabiner LR (1989) A tutorial on hidden Markov models and selected applications in speech recognition. *Proc IEEE* 77(2):257-286.
15. Fink GA (2013) Mustererkennung mit Markov-Modellen: Theorie - Praxis - Anwendungsgebiete. (Vieweg+Teubner, Stuttgart). (English version available at Springer, London)
16. van de Meent J, Bronson JE, Wiggins CH & Gonzalez Jr. RL (2014) Empirical Bayes Methods Enable Advanced Population-Level Analyses of Single-Molecule FRET Experiments. *Biophys J* 106(6):1327-1337.
17. Chen Y, Shen K, Shan S & Kou SC (2016) Analyzing Single-Molecule Protein Transportation Experiments via Hierarchical Hidden Markov Models. *J Am Stat Assoc* doi:10.1080/01621459.2016.1140050, published online: Feb 2.
18. Keller BG, Kobitski A, Jäschke A, Nienhaus GU & Noé F (2014) Complex RNA Folding Kinetics Revealed by Single-Molecule FRET and Hidden Markov Models. *J Am Chem Soc* 136(12):4534-4543.
19. Blanco MR et al. (2015) Single Molecule Cluster Analysis dissects splicing pathway conformational dynamics. *Nat Meth* 12(11):1077-1084.
20. Lee NK et al. (2005) Accurate FRET Measurements within Single Diffusing Biomolecules Using Alternating-Laser Excitation. *Biophys J* 88(4):2939-2953.
21. Kelly D, Dillingham M, Hudson A & Wiesner K (2012) A New Method for Inferring Hidden Markov Models from Noisy Time Sequences. *PLoS ONE* 7(1):e29703.
22. Schwarz G (1978) Estimating the Dimension of a Model. *Ann Statist* 6(2):461-464.
23. Bruno WJ, Yang J & Pearson JE (2005) Using independent open-to-closed transitions to simplify aggregated Markov models of ion channel gating kinetics. *Proc Natl Acad Sci USA* 102(18):6326-6331.
24. Flomenbom O, Klafter J & Szabo A (2005) What Can One Learn from Two-State Single-Molecule Trajectories?. *Biophys J* 88(6):3780-3783.
25. Amann CP, Schmiedl T & Seifert U (2010) Communications: Can one identify nonequilibrium in a three-state system by analyzing two-state trajectories?. *J Chem Phys* 132(4):041102.
26. Myung IJ (2003) Tutorial on maximum likelihood estimation. *J Math Psychol* 47(1):90-100.
27. Giudici P, Rydén T & Vandekerkhove P (2000) Likelihood-Ratio Tests for Hidden Markov Models. *Biometrics* 56(3):742-747.
28. Panaretou B et al. (1998) ATP binding and hydrolysis are essential to the function of the Hsp90 molecular chaperone in vivo. *EMBO J* 17(16):4829-4836.
29. Howard J (2001) Mechanics of Motor Proteins and the Cytoskeleton. (Sinauer Associates, Sunderland).
30. Hill TL (1989) Free energy transduction and biochemical cycle kinetics. (Springer, New York).
31. Phillips RC, George P & Rutman RJ (1969) Thermodynamic Data for the Hydrolysis of Adenosine Triphosphate as a Function of pH, Mg2+ Ion Concentration, and Ionic Strength. *J Biol Chem* 244(12):3330-3342.
32. Suzuki T, Tanaka K, Wakabayashi C, Saita E & Yoshida M (2014) Chemomechanical coupling of human mitochondrial F1-ATPase motor. *Nat Chem Biol* 10(11):930-936.
33. Panaretou B et al. (2002) Activation of the ATPase Activity of Hsp90 by the Stress-Regulated Cochaperone Aha1. *Mol Cell* 10(6):1307-1318.
34. Pearl LH (2016) The HSP90 molecular chaperone – an enigmatic ATPase. *Biopolymers* doi:10.1002/bip.22835, published online: Mar 17.
35. Liu Y, Park J, Dahmen KA, Chemla YR & Ha T (2010) A Comparative Study of Multivariate and Univariate Hidden Markov Modelings in Time-Binned Single-Molecule FRET Data Analysis. *J Phys Chem B* 114(16):5386-5403.
36. Hessling M, Richter K & Buchner J (2009) Dissection of the ATP-induced conformational cycle of the molecular chaperone Hsp90. *Nat Struct Mol Biol* 16(3):287-293.
37. Jahn M et al. (2014) The charged linker of the molecular chaperone Hsp90 modulates domain contacts and biological function. *Proc Natl Acad Sci USA* 111(50):17881-17886.



\twocolumn[{
{\Huge\textbf{Supporting Information}}
\newline

{\LARGE \textbf{Schmid et al.}}
\vspace{0.4cm}
}]

\subsection*{SI Methods}

\textbf{General implementation of semi-ensemble HMM.} 
In the following, we include all formulae required for the implementation of semi-ensemble HMM as demonstrated herein. For more general introductions to HMM, please refer to the respective literature (14,15). \\
Forward-Backward, Baum-Welch and Viterbi algorithms were implemented for continuous observables and multiple dimensions. Numerical underflow or overflow is prevented by logarithmic renormalization. Recursive calculations are sped up by multi-threading (processing several time-traces in parallel). All software was written in IgorPro v6.3 (Wavemetrics) and calculations were run on an iMac (Apple, 2014, 2.9 GHz Intel i5 processor, 16GB RAM) or a comparable Windows PC. A typical optimization (4 states, >100 traces) took less than an hour.
\newline

\textbf{HMM conventions and parameters.} 
The indices $i, j$ denote states. $t$ are discrete time steps and $T$ is the total time of a trajectory. $O$ is the set of observables and $\bm{x}_t$ is a specific observable at time $t$ in $d$ dimensions (herein $d=2$ for donor and acceptor fluorescence).
The complete set of parameters, $\lambda(\bm\pi, \bf A, B)$, consists of:  \vspace{-0.4cm}

\begin{align*}
\pi_i &\overset{\scriptscriptstyle\wedge}{=} \text{start probabilites}\\
a_{ij} &\overset{\scriptscriptstyle\wedge}{=} \text{transition probabilities}\\
b_i(\bm{\mu}_i, V_{i}) &\overset{\scriptscriptstyle\wedge}{=} \text{Gaussian emission probability densities} 
\end{align*}

with means $\bm\mu_i$ and covariance matrix $V_i$ , both in $d$ dimensions. TIRF data is appropriately described by Gaussian emissions (in place of Poissonian), because each time bin contains much more than ten photons including noise.
$b_i(\bm{x}_t)$ denotes the emission density value for a specific observable value at a given time t (scalar, may be >1).
\newline

\textbf{Implementation of the forward-backward algorithm.} 
The forward and backward variables, $\alpha$ and $\beta$, are auxiliary probabilities prerequisite for the Baum-Welch algorithm below. 

\begin{align*}
	\text{initiation:}&& 	\alpha_{t=1}(i) &= \pi_i \, b_i(\bm{x}_{t=1})	\\
	\text{recursion:}&& 	\alpha_{t+1}(i) &= \sum_j \left[ \alpha_t(j)\,a_{ji}\right]\, b_i(\bm{x}_{t+1})	\\
	\text{termination:}&& P(O|\lambda) &= \sum_{i} \alpha_T(i)	\\
	\text{initiation:}&& 	\beta_T(i) &= 1	\\
	\text{recursion:}&&	 \beta_t(i) &= \sum_j a_{ij} \,b_j(\bm{x}_{t+1})\,\beta_{t+1}(j)	\\
	\text{termination:}&&  P(O|\lambda) &= \sum_{i} \pi_i \,b_i(\bm{x}_1)\,\beta_{t=1}(i)
\end{align*}

$P(O|\lambda)$ is called the production probability of the data given the model. It is equivalent to the likelihood of the model given the data $\mathcal{L}(\lambda|O)$. 
\newline

\textbf{Implementation of the Baum-Welch algorithm.} 
The basis for calculating the updated parameters are $\gamma_t(i)$ and $\gamma_t(i,j)$, the respective probabilities for a given state or transition at a given time point. 

\begin{align*}
	\gamma_{t}(i) &= \alpha_t(i) \,\beta_t(i) / P(O|\lambda)	\\
	\gamma_{t}(i,j) &= \alpha_t(i) \,a_{ij} \,b_j(\bm{x}_{t+1}) \,\beta_{t+1}(j) / P(O|\lambda)
\end{align*}

The parameter update equations are: \vspace{-0.5cm}

\begin{align*}	
	\hat{\pi}_i &= \gamma_{t=1}(i)\\
	\hat{a}_{ij} &= \sum\nolimits_{t=1}^{T-1} \left[ \gamma_t(i,j) \right] / \sum\nolimits_{t=1}^{T-1} \gamma_t(i)	\\
	\hat{\bm{\mu}}_{i} &= \sum\nolimits_{t=1}^{T} \left[\gamma_t(i) \,\bm{x}_t \right]/ \sum\nolimits_{t=1}^{T} \gamma_t(i) 	\\
	\hat{V}_{i} &=  \sum\nolimits_{t=1}^{T} \left[\gamma_t(i) \,\bm{x}_t \,  {\bm{x}_t}^\mathsf{T}       \right]/ \sum\nolimits_{t=1}^{T}\left[ \gamma_t(i) \right]  - \bm{\mu}_i \, {\bm{\mu}_i}^\mathsf{T}  
\end{align*}

The ensemble parameters $\Pi_i$, $\mathcal{A}_{ij}$ are updated based on all $N$ time-traces: \vspace{-0.4cm}

\begin{align*}	
	\hat{\Pi}_i &=\sum\nolimits_{n=1}^N \left[^n\pi_{i} \right]/ N	\\
	\hat{\mathcal{A}}_{ij} &= \sum\nolimits_{n=1}^{N} \Big[ 	\sum\nolimits_{t=1}^{T-1} \left[ 	^n\gamma_t(i,j) \right] 	\Big]	/ 	\sum\nolimits_{n=1}^{N} \Big[ 	\sum\nolimits_{t=1}^{T-1} \left[ 	^n\gamma_t(i) \right] 	\Big]	
\end{align*}

\vspace{0.3cm}
\textbf{Implementation of the FRET-constraint.} 
In a true FRET time-trace, the total fluorescence $I_{tot}$ (i.e. the sum of the corrected acceptor and donor signals, ${^A{x}_t}$ and ${^D{x}_t}$, with respective means $^A\mu_i$ and $^D\mu_i$) must remain constant in each state: \vspace{-0.4cm}

\begin{align*}	
	\langle I_{tot} \rangle = \sum\nolimits_{t=1}^{T}  \big[ {^A{x}_t} + {^D{x}_t} \big] / T =  {^A\mu_i} + {^D\mu_i} &=   const. \quad \forall \, i	
\end{align*}

This physical constraint is introduced into Baum's optimization formalism using Lagrange multipliers.
Because the resulting update equations for Gaussian distributions are coupled in $\bm{\mu}_i$ and $V_i$,
 we exploit the fact that the difference between Gaussian and Poissonian means is negligible for TIRF signals.
Therefore, the update equation for constrained Poissonian distributions (35) can be utilized to optimize $\bm{\mu}_i$ :	 \vspace{-0.4cm}

\begin{align*}	
	\hat{\bm{\mu}}_{i} &= \langle I_{tot} \rangle \cdot \sum\nolimits_{t=1}^{T} \left[\gamma_t(i) \,\bm{x}_t \right]  /   \sum\nolimits_{t=1}^{T} \left[ \gamma_t(i) \cdot (^A{x}_t + {^D}{x}_t)	\right]
\end{align*}

\vspace{0.3cm}
\textbf{Implementation of the Viterbi algorithm.}
The most probable state sequence $s^*$ with maximal production probability $P^*(O|\lambda)$ is deduced from the $\delta$ and $\psi$ variables. \vspace{-0.4cm}

\begin{align*}
	\text{initiation:}&&\delta_{t=1}(i) &= \pi_i \, b_i(\bm{x}_{t=1})	\\
	\text{recursion:}&&\delta_{t+1}(j) &= \max_i [\delta_t(i)a_{ij}]b_j(\bm{x}_{t+1})	\\
	\text{termination:}&&P^*(O|\lambda) &= P(O,s^*|\lambda) =  \max_i \delta_T(i) \\
				   &&s_T^* &= \underset{j}{\operatorname{argmax}} [\delta_T(j)]		\\
	\text{initiation:}&&\psi_{t=1}(i) &= 0	\\
	\text{recursion:}&&\psi_{t+1}(j) &= \underset{i}{\operatorname{argmax}}[\delta_t(i)a_{ij}]	\\
	\text{back-tracking:}&&s_t^* &= \psi_{t+1}(s^*_{t+1})
\end{align*}

\vspace{0.3cm}
\textbf{The Bayesian information criterion (BIC).}
The BIC (22) 
 selects for a model that describes the data well, while keeping the model complexity moderate. This is achieved by balancing the likelihood $\mathcal{L}(\lambda|O)$ against the number of free parameters $k$, with $n$, the number of data points: 

\begin{equation*}
\text{BIC} = -2\cdot \ln(\mathcal{L}) + k\cdot \ln(n)
\end{equation*}
Its applicability in the context of SMACKS was confirmed using synthetic data of known input models (see below).
\newline

\textbf{Simulations.}
Discrete \textit{state sequences} were obtained by a Monte-Carlo simulation based on a given transition matrix: photo-bleaching was included by exponential trace length distributions. For comparison with experimental data, corresponding minimal trace lengths were used (typically 30 data points).\\
As in the experiment, \textit{synthetic data} contained Gaussian noise ($\upsigma$=0.3*signal), random offsets ($\pm$0.2*signal), degenerate FRET efficiencies (two low / two high), a sampling rate of 5Hz and a bleach rate of 0.03Hz. See example data in Figure S2.
\newline

\textbf{Confidence Interval}
For each transition probability $a_{ij}$, the parameter space around the maximum likelihood estimator (MLE) is scanned while keeping the remaining parameters fixed at the MLE. The modified models $\lambda'$ are compared to the MLE models $\lambda_\textrm{MLE}$ by successive likelihood ratio ($LR$) tests (3,27):\vspace{-0.4cm}

\begin{align*}
LR &= 2 \left( \, \ln[\mathcal{L}(\lambda_{\textrm{MLE}} | O)] - \ln[ \mathcal{L}(\lambda' | O)] \,\right) \\
a_{ij}^{\textrm{CB}} &=  a'_{ij}  \quad : \quad  LR = \chi^2_{0.95\textrm{,df}=1} = 3.841
\end{align*}

The 95\% confidence bound (CB) is reached where $LR$ crosses the respective significance level for one degree of freedom (df).
\newline

\textbf{Sample Preparation.}
\textit{Hsp90:} Yeast Hsp90 dimers (UniProtKB: P02829) supplied with a C-terminal zipper motif were used to avoid dissociation at low concentrations (13). 
Previously published cysteine positions (36) allowed for specific labeling with donor (61C) or acceptor (385C) fluorophores (see below). Both constructs were cloned into a pET28b vector (Novagen, Merck Biosciences). They include an N-terminal His-tag followed by a SUMO-domain for later cleavage. The QuickChange Lightning kit (Agilent) was used to insert an Avitag for specific \textit{in vivo} biotinylation at the C-terminus of the acceptor construct. \textit{E.coli} BL21star cells (Invitrogen) were co-transformed with pET28b and pBirAcm (Avidity) by electroporation (Peqlab) and expressed according to Avidity's \textit{in vivo} biotinylation protocol. The donor construct was expressed in \textit{E.coli} BL21(DE3)cod+ (Stratagene) for 3h at 37°C after induction with 1mM IPTG at OD\textsubscript{600}=0.7 in LB\textsubscript{Kana}. A cell disruptor (Constant Systems Ltd.) was used for lysis. Proteins were purified as published (37) (Ni-NTA, tag cleavage, anion exchange, size exclusion). 95\% purity was confirmed by SDS-PAGE.
Fluorescent labels (Atto550-, Atto647N-maleimide) were purchased from Atto-tec and coupled to cysteins according to the supplied protocol. Hetero-dimers (acceptor+donor) were obtained by 20min incubation of 1$\upmu$M donor, 0.1$\upmu$M biotinylated acceptor homo-dimers and 2mM ADP in measurement buffer (40mM Hepes, 150mM KCl, 10mM MgCl$_2$) at 47°C. In this way, predominantly biotinylated \textit{hetero}-dimers bind to the neutravidin (Thermo Fisher) coated fluid chamber (see below). (Residual homo-dimers will show a specific smFRET signal and are excluded from analysis.)\\
\textit{Holliday junction:} DNA-oligos similar to (8) 
with fluorophores Atto488, Atto550 and Atto647N (Atto-tec) attached were purchased from IBA GmbH. A mixture of 100nM of each DNA-oligo in Tris-buffer (5mM Tris, 5mM NaCl, 20mM MgCl\textsubscript{2}, pH 7.5) was heated to 90°C for 10 min and cooled down to 20°C (1°C/min) in a thermocycler (Peqlab). Holliday junctions were measured in Tris-buffer (5mM Tris, 5mM NaCl, 500mM MgCl\textsubscript{2}) including 0.1\% glucose, 10 U/ml glucose oxidase (\textit{Aspergillus niger}), 100 U/ml catalase (bovine liver, Calbiochem), 2mM Trolox.\\
If not stated differently, all chemicals were purchased from Sigma Aldrich.
\newline

\textbf{Single-Molecule Setup \& Measurements.}
TIRF setup: An objective-type TIRF setup was built to measure smFRET. Green and red excitation lasers (532nm, Compass 215M, Coherent and 635nm, Lasiris, Stocker Yale) were aligned, expanded and focused onto the back-focal plane of an apochromat TIRF 100x objective (Nikon, NA=1.49). The collected fluorescence is separated from excitation light by notch filters. Off-axis beams are removed by an optical slit and achromatic slit lenses (achromatic doublets, Qioptiq). Finally donor and acceptor fluorescence is split, spectrally filtered again and individually focused side by side onto the EMCCD (iXonUltra, Andor) by best form silica lenses (Qioptiq). Translation stages (Newport) were used to fine-tune the lens positions. Dichroic mirrors and filters were purchased from AHF analysentechnik AG. Optical mounts, lenses, mirrors and further components were purchased from Thorlabs unless stated differently.
Measurements were performed in a PEGylated fluid chamber built from two coverslips and Nescofilm by compression at 70°C. To avoid auto-fluorescence background in the red detection channel a silica coverslip was used (Spectrosil2000, Heraeus, manufactured by UQG Ltd.). The refractive index change (glass: 1.52, silica: 1.46) was adjusted for geometrically. Image acquisition and optical shutters were synchronized by a custom-built electronic circuit and a master trigger to achieve 100ms exposure to each ALEX channel (alternating laser excitation).
\newline

\textbf{Selection \& Correction of smFRET Time Traces.}
We exclude incomplete FRET pairs and photo-physical artifacts, such as blinking, by using ALEX for data acquisition. Selection criteria for single molecule time traces are flat plateaus in all three fluorescence channels (donor and acceptor fluorescence after donor excitation and acceptor fluorescence after acceptor excitation), as well as single step bleaching. Hence, a terminating low FRET region (before photo-bleaching) is no longer rejected. Fluorescence time traces are corrected for background offsets, leakage, direct excitation and the gamma factor (20). 
\newline

\subsection*{SI Notes}

\textbf{Note 1: Hierarchical search for simplified models.}
Bruno et al. (23) describe a procedure to deduce the simplest, plausible reaction schemes, from data with multiple open and closed conformations, by comparing models of the canonical “MIR”-form (manifest interconductance rank). We consider a 4-state model with 2 open (o) and 2 closed (c) states (N\textsubscript{o}=N\textsubscript{c}=2), as previously determined by BIC and the bi-exponential dwell time distributions.
First, the interconductance rank (i.e. the number of independent o-c links) is determined. To this end, MIR-form models of rank 1 (linear o-o-c-c) and rank 2 (cyclic -o-o-c-c-) are compared in a likelihood ratio ($LR$) test (likelihood of rank $x$ model, $L_{Rx}$):
\begin{align*}
LR = 2\cdot[ \ln(L_{R2}) -  \ln(L_{R1}) ]  \;	
    \begin{cases} 
    ≤ \chi^2_{0.95, df=2} \quad \Rightarrow \textrm{rank 1}\\ 
    > \chi^2_{0.95, df=2} \quad \Rightarrow \textrm{rank 2}
    \end{cases}	
\end{align*}
The null hypothesis (rank 1 model) is rejected if the likelihood ratio exceeds the 95\% confidence interval given by the $\chi^2$-distribution for 2 degrees of freedom ($df$). (One missing link equals a difference of two transitions.)
Second, the number of links $N_l$ within this rank R is determined by comparing different schemes by BIC. The number of mathematically identifiable links is limited:
\newpage
\begin{align*}
 N_l≤R(N_o+N_c-R) 
\end{align*}
Models with the same rank and the same number of links are mathematically equivalent and cannot be discerned without further experimental data. For Hsp90, we find a cyclic -o-o-c-c- model in the presence of ATP and linear o-o-c-c models for apo, ADP or AMP-PNP conditions.
\newline

\textbf{Note 2: Expected value of observed energy coupling.}
Let $\langle N_{ij}^{obs} \rangle$ be the expected number of observations for a given transition, in a dataset with $N_{tot}$ data points. If it is smaller than one, the specific transition cannot be resolved. In this case, $f_{ij}^{\textrm{lost}}$ denotes the factor that is actually lost and the expected value of the observed free energy change $\langle \Delta G_{obs}\rangle$ in units of kT is given by:
\begin{align*}
\langle \Delta G_{obs} \rangle =  - \sum_{\substack{\forall i ≠ j \\ \textrm{ (cycl.)}}} \ln\left[ \frac{a_{ij} / f_{ij}^{\textrm lost}}{a_{ji} / f_{ji}^{\textrm lost}} \right]	
\end{align*}
with: 
\begin{align*}
 f_{ij}^{\textrm lost} =
    \begin{cases} 
    \langle N_{ij}^{obs} \rangle  = N_{tot}  \cdot \pi_i \cdot a_{ij} 	\quad &\forall \; \langle N_{ij}^{obs} \rangle < 1 \\ 
    \; 1		\quad &\forall \; \langle N_{ij}^{obs} \rangle ≥ 1
    \end{cases}	
\end{align*}

\onecolumn

\begin{figure}[ht] 
   \centering
   \includegraphics[width=11.4cm]{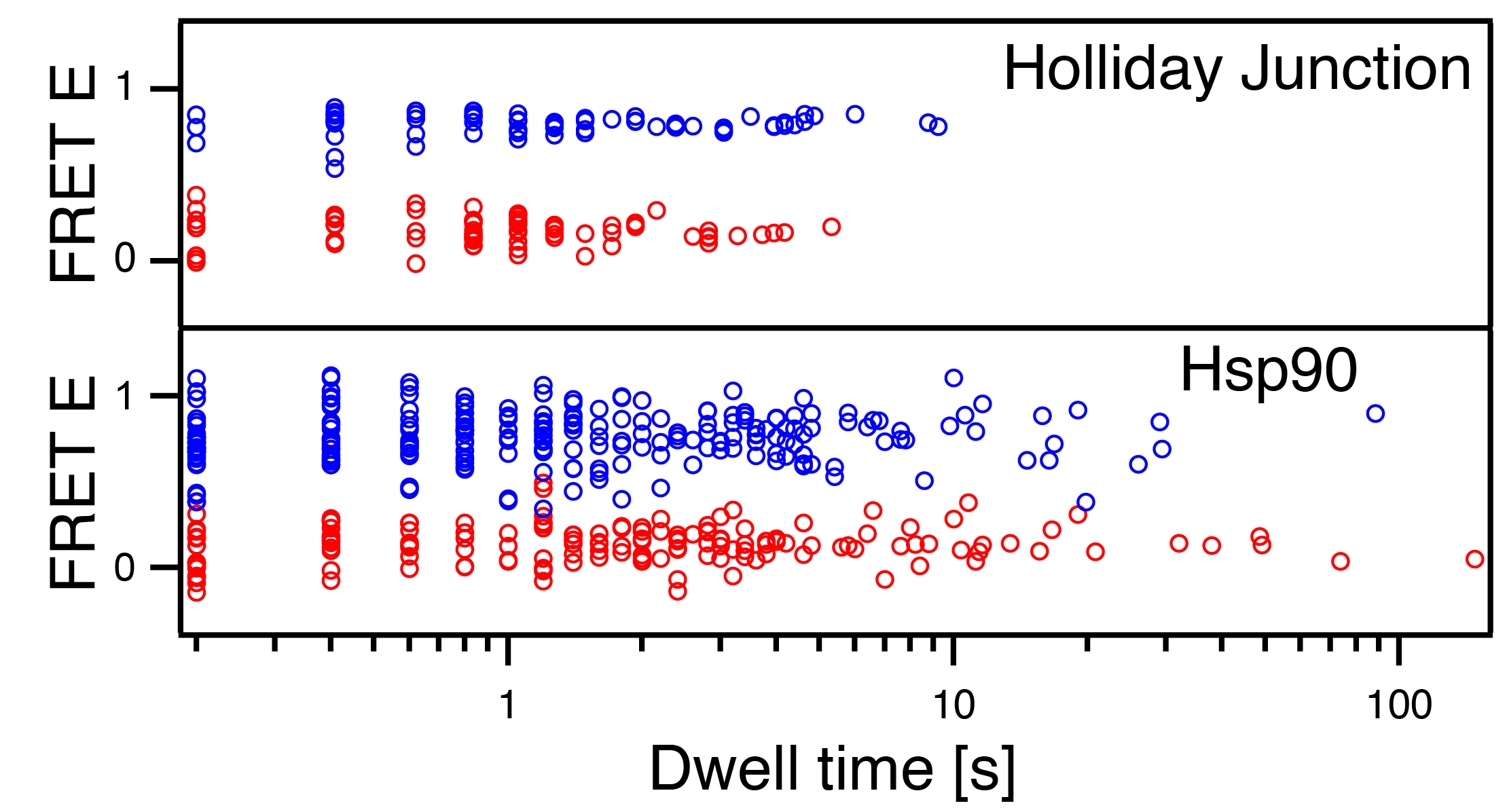} 
   \caption{FRET efficiency vs. dwell time plots for a Holliday junction or Hsp90 (163 molecules) as indicated. Low FRET dwells in red; high FRET dwells in blue. No correlation is visible in either plot. In contrast to the Holliday junction, Hsp90’s conformational changes occur over a much broader time range.}
    \vspace{1.5cm}   

   \includegraphics[width=8.7cm]{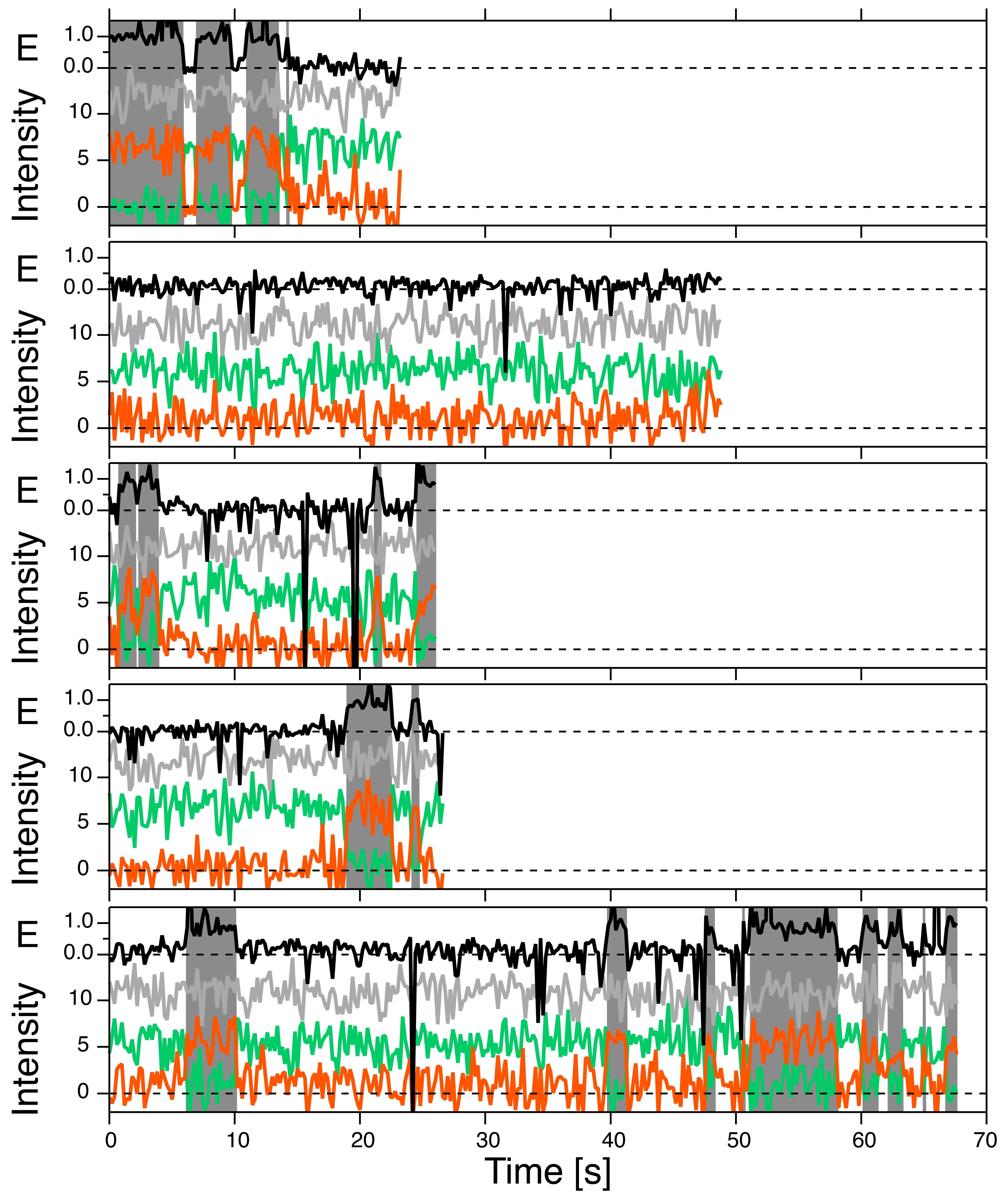} 
   \caption{Representative synthetic data generated by a 4-state model with 2x2 degenerate FRET efficiencies as described in the Online Methods. Fluorescence intensity of the donor (green), acceptor (orange) and acceptor after direct excitation (gray). Calculated FRET efficiency (E) in black and Viterbi path as overlays (two states: high FRET gray, low FRET white). Static traces (second trace) occur as a consequence of fast and slow rates together with a finite observation time due to photo-bleaching. FRET efficiency spikes occur in the simulations as well as in the experiment. In contrast to 1D HMM based on FRET, they have no effect on fluorescence based 2D HMM as can be seen from the Viterbi paths.}
\end{figure}

\begin{figure}[ht] 
   \centering
   \includegraphics[width=11.4cm]{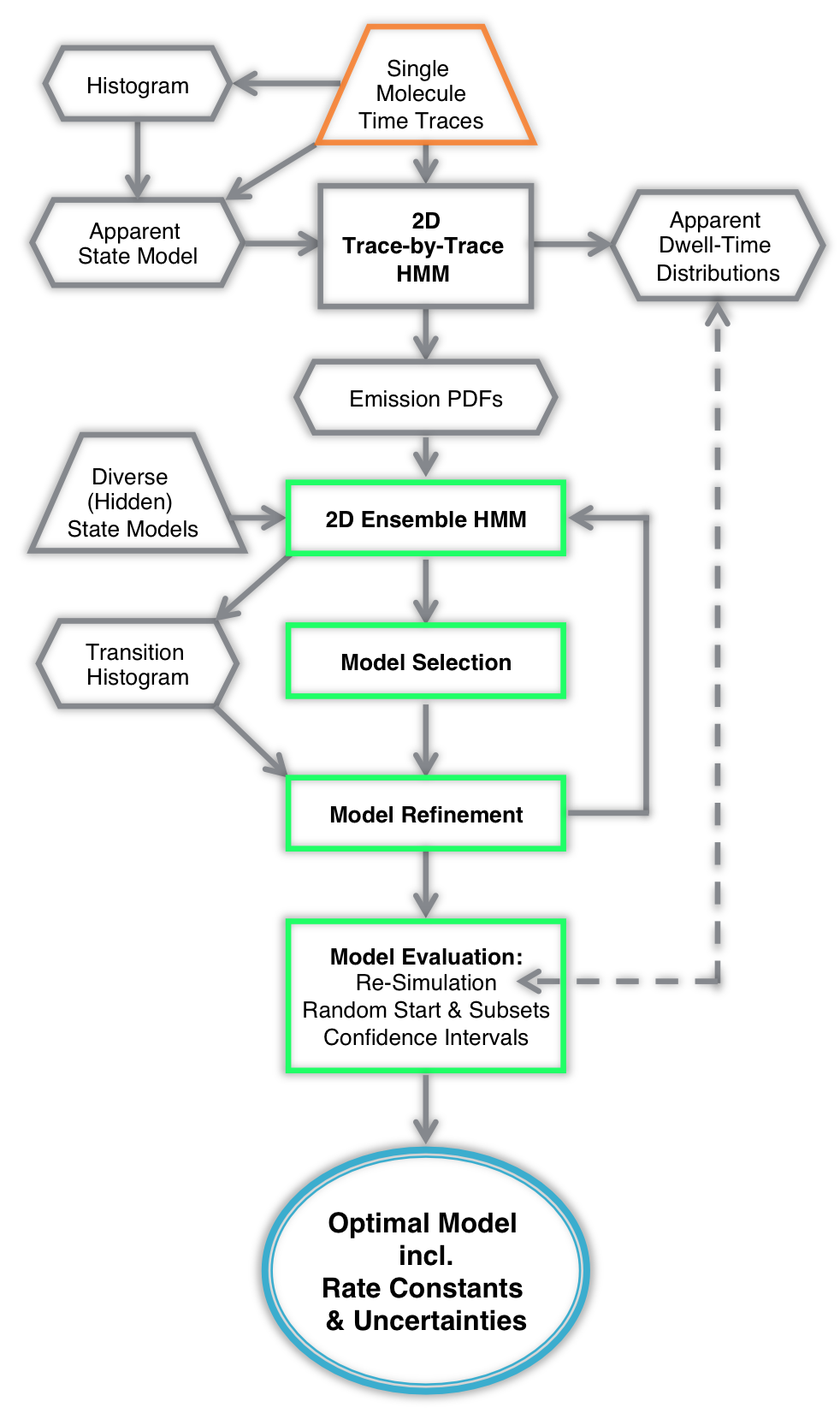} 
   \caption{SMACKS as a flow chart.}
\end{figure}

\end{document}